# Dual-comb femtosecond solid-state laser with inherent polarization-multiplexing

Maciej Kowalczyk[1,*], Łukasz Sterczewski[1], Xuzhao Zhang[2,3], Valentin Petrov[4], Zhengping Wang[2], Jarosław Sotor[1]

[1]Laser & Fiber Electronics Group, Faculty of Electronics, Wroclaw University of Science and Technology, Wybrzeże Wyspiańskiego 27, 50–370 Wroclaw, Poland
[2]State Key Laboratory of Crystal Materials, Shandong University, 250100 Jinan, China
[3]Center of Nanoelectronics, School of Microelectronics, Shandong University, 250100 Jinan, China
[4]Max Born Institute for Nonlinear Optics and Ultrafast Spectroscopy, Max-Born-Str. 2a, 12489 Berlin, Germany
*Email: m.kowalczyk@pwr.edu.pl

**Abstract**

Dual-comb spectroscopy is a rapidly developing technique enabling ultraprecise broadband optical diagnostics of atoms and molecules. This powerful tool typically requires two phase-locked femtosecond lasers, yet it has been shown that it can be realized without any stabilization if the combs are generated from a single laser cavity. Still, unavoidable intrinsic relative phase-fluctuations always set a limit on the precision of any spectroscopic measurements, hitherto limiting the applicability of bulk dual-comb lasers for mode-resolved studies. Here, we demonstrate a versatile concept for low-noise dual-comb generation from a single-cavity femtosecond solid-state laser based on intrinsic polarization-multiplexing inside an optically anisotropic gain crystal. Due to intracavity spatial separation of the orthogonally-polarized beams, two sub-100 fs pulse trains are simultaneously generated from a 1.05 µm Yb:CNGS oscillator with a repetition rate difference of 4.7 kHz. The laser exhibits the lowest relative noise ever demonstrated for a bulk dual-comb source, supporting free-running mode-resolved spectroscopic measurements over a second. Moreover, the developed dual-comb generation technique can be applied to any solid-state laser exploiting a birefringent active crystal, paving the way towards a new class of highly-coherent, single-cavity, dual-comb laser sources operating in various spectral regions.

## 1. Introduction

Dual-comb spectroscopy (DCS) has played a pivotal role in modern optical metrology by enabling to acquire simultaneously broadband and high resolution optical spectra at unprecedentedly high speeds [1,2]. Unlike Fourier Transform spectroscopy, DCS does not require any moving parts for interferometry, while in contrast to single-frequency diode lasers conventionally used in solid-state spectrometers, the obtainable spectral coverage in DCS routinely reaches multiple terahertz. The concept of DCS relies on an interaction between two lasers with a unique spectral pattern of narrow, equidistant and mutually locked lines known as optical frequency combs (OFCs). A mismatch of their line spacings enables multi-heterodyne beating on a square-law photodetector resulting in a transfer of the optical comb structure to the radio-frequency (RF) domain. Although a pair of OFCs can be generated from a single continuous-wave (CW) laser through dual electro-optical modulation [3] or nonlinear Kerr interaction inside a pair of microresonators [4], arguably the most conventional DCS implementation employs a pair of spectrally-overlapping, yet independently running mode-locked lasers. Nevertheless, an attempt to perform precise free-running DCS measurements in this configuration without any synchronization loops is rather unsuccessful. This relates to uncorrelated fluctuations of the combs' line spacings ($f_{rep}$) and their starting points defined by a carrier-envelope-offset frequency ($f_{ceo}$), which induce frequency instabilities of the DCS beating signal referred to as the interferogram (IFG). A loss of the discrete character of the RF comb in many cases makes precise DCS measurements impossible. Although frequency instabilities are inherent to all lasers, the IFG phase noise can be strongly reduced via active phase-locking, which allows to achieve long-term coherence sufficient for spectroscopic measurements [5,6]. Unfortunately, this requires simultaneous synchronization of $\Delta f_{rep}$ and $\Delta f_{ceo}$ between the two lasers, which, despite several years since its first demonstration, still appears as a challenging task.

An alternative concept stems from the idea to generate two mode-locked pulse trains with different repetition rates from a single laser cavity. The strength of this technique lies in the intrinsic common-mode nature of phase fluctuations for the two combs, which renders their high mutual coherence and virtually eliminates the need for phase-locking electronics. Because the beams share the same cavity, most of the relative phase noise originating from thermal or mechanical perturbations cancels out. Consequently, as long as the uncorrelated phase excursions between the two OFCs remain minor, the combs exhibit sufficient relative stability to enable free-running DCS over second time scales.

Historically, the first single-cavity dual-comb mode-locked laser was a 1.55 µm fiber oscillator operating in a bidirectional configuration [7,8]. Subsequently, similar fiber sources have been demonstrated in various spectral ranges with dual-comb generation based on polarization-multiplexing [9–11] or multiwavelength emission [12–15]. Despite these advances, the main drawback of soliton mode-locked lasers based on all-fiber technology is the limitation of their output power due to excessive nonlinearities, which typically reduces the achievable peak power below the kilowatt level. This in turn prevents their efficient nonlinear frequency conversion to other spectral ranges (e.g. mid-IR), where many important chemical species have their strongest absorption features. Although it has been recently shown that the pulse energy can be enhanced by operating a fiber laser in the normal dispersion regime [16], it always comes at the expense of chirping the pulse to the picosecond level. The power limitation can be in principle mitigated by exploiting solid-state lasers; often with superior noise properties over their fiberized counterparts [17]. To date, however, only few examples of bulk dual-comb lasers can be found in the literature.

Although dual-wavelength (time-synchronized) Ti:sapphire femtosecond oscillators were demonstrated in the early 1990s [18], the first single-cavity dual-comb bulk laser was reported more than 20 years later in 2015. A semiconductor mode-locked integrated external-cavity surface emitting laser (MIXSEL) generated 18 ps long pulses at ~0.97 µm [19]. The simultaneous emission of two mutually coherent combs relied on the presence of an additional birefringent medium in the laser cavity. A pair of orthogonally-polarized beams experienced a double refraction inside the calcium carbonate crystal, which resulted in their spatial separation and hence different intracavity path lengths. The same concept was later also applied to a picosecond Nd:YAG laser [20]. From an application perspective, MIXSEL dual-comb lasers have already proven their usefulness in molecular spectroscopy of gaseous acetylene at 1.03 µm [21] and water vapor at 0.97 µm [22]. Another interesting dual-comb architecture has been demonstrated in a monolithic laser comprising a single highly-reflective (HR)-coated Yb:KGW crystal [23]. Polarization-splitting originating from the gain properties of this biaxial active crystal enabled generation of two mode-locked pulse trains with repetition rates of ~25 GHz and ~1 ps pulse duration at 1.05 µm. Unfortunately, this configuration did not allow to control the energy balance between the beams. Therefore, the dual-comb regime could only be achieved for the special case of equal gain for the two polarizations, yet with an achievable bandwidth limited to a few nanometers. In 2017, a bidirectional mode-locked Ti:sapphire oscillator based on a ring-cavity was demonstrated [24]. The generation of two counterpropagating combs with ultrashort 12-fs pulse duration, centered around 0.84 µm was attributed to the nonlinear self-steepening phenomenon. Very recently, a new scheme for dual-comb generation has been implemented in ytterbium thin-disk lasers emitting at ~1.03 µm. The concept exploited spatial separation of two beams enabled by non-common cavity end mirrors [25,26] and/or polarization-splitting [27]. The unprecedented output power levels of several watts achieved for pulse durations on the order of 300 fs promise effective nonlinear frequency conversion of the dual-comb structure towards the mid-IR or UV domains. Nevertheless, neither of these two systems can be termed as single-cavity, because in both cases the intracavity beams did not share all the resonator mirrors, which is expected to compromise the mutual coherence between the generated OFCs.

In this paper, we present a novel concept for dual-comb generation from a single-cavity mode-locked solid-state laser based on an optically anisotropic gain crystal. It relies on simultaneous generation of two spatially-separated beams and it builds on the idea of polarization-multiplexing originating from the double refraction phenomenon described in [19]. In contrast to the latter work, our scheme does not require any additional components to be added to a standard laser cavity, as it is based on the inherent birefringence of the employed gain medium. If the oscillating laser beam does not propagate along the optical axis of a uniaxial active crystal, its path will vary depending if it is polarized normal to the principal plane (an ordinary (*o*)-ray) or in the principal plane (extraordinary (*e*)-ray). However, if the propagation is also not at 90° to the optical axis, then the two eigen-polarizations will experience a spatial offset. It is possible to achieve simultaneous lasing at both polarizations, if the corresponding rays both overlap with the pump beam inside the active medium. Ideally, this can be realized if the angles of incidence (AOI) on the crystal are different for the *o*/*e*-beams, so that the birefringent spatial walk-off is compensated and both Poynting vectors overlap being collinear with the pump. Consequently, the orthogonally-polarized laser beams will follow different intracavity paths (providing non-zero $\Delta f_{rep}$ in the mode-locked regime), yet they will stay spatially-combined inside the gain medium due to the double refraction phenomenon. Note, that the feasibility to control the relative overlap of the beams permits to counterbalance the possible anisotropy of the gain in a birefringent active crystal. Due to the single-cavity design with minimized number of components and a single pump source, one can employ the presented technique to achieve remarkably low-noise dual-comb generation.

We demonstrate experimental realization of the described concept in an Yb:Ca$_3$NbGa$_3$Si$_2$O$_{14}$ (calcium niobium gallium silicate; Yb:CNGS) mode-locked laser. The developed oscillator simultaneously emits two spatially separated pulse trains at 1.05 µm with a repetition rate difference of 4.73 kHz around 78.3 MHz. The pulse durations of the combs amount to 88 and 93 fs, making this source the first sub-100 fs dual-comb single-cavity ytterbium laser. The performed characterization of the mutual coherence between the combs indicates sub-kilohertz relative linewidth over 10 IFG periods, which permits free-running mode-resolved DCS over milliseconds simultaneously supporting further enhancement of the measurement precision to second timescales by computational phase correction.

## 2. Experimental setup

The experimental setup of our dual-comb spectrometer is shown in Figure 1. The core of the system is a mode-locked Yb:CNGS laser with a cavity design based on our previous works [28,29]. As a gain medium we used an antireflective (AR)-coated 3 at.% Yb:CNGS crystal with a length of 4 mm and an aperture of 3 × 3 mm$^2$. It was pumped by a semiconductor laser diode delivering up to 1 W of 979.4 nm radiation at a maximum driving current of 1.7 A (3SP Technologies). The polarization-maintaining (PM) single-mode-fiber-coupled diode provided diffraction-limited pump beam quality and exhibited intrinsically low-noise operation compared to powerful multimode diodes [30,31]. The radiation from the diode was reimaged inside the gain medium with a resulting focal radius of 25 µm. Due to the modest pump power level used here, active crystal cooling was not necessary. The pump polarization was aligned as vertical to exploit the higher absorption of Yb:CNGS for σ-polarization [32]. At normal incidence, both the pump and the *o*-polarized laser beam are collinear outside the crystal. The gain of Yb:CNGS is also higher for the σ-polarization, but as already mentioned, the adjustable spatial overlap makes the present architecture much more flexible compared to [23] with respect to equalizing the net gain for the two principal polarizations. For mode-locked operation we employed a semiconductor saturable absorber mirror (SESAM) with a modulation depth of 2.6% and a relaxation time of 500 fs

(Batop). Anomalous group-delay dispersion (GDD) required for soliton mode-locking was introduced with three Gires-Tournois Interferometer (GTI) mirrors with a net round-trip (i.e. double-bounce) GDD amounting to −1800 fs$^2$ (Layertec). The output coupler (OC) had a transmittance of 1%.

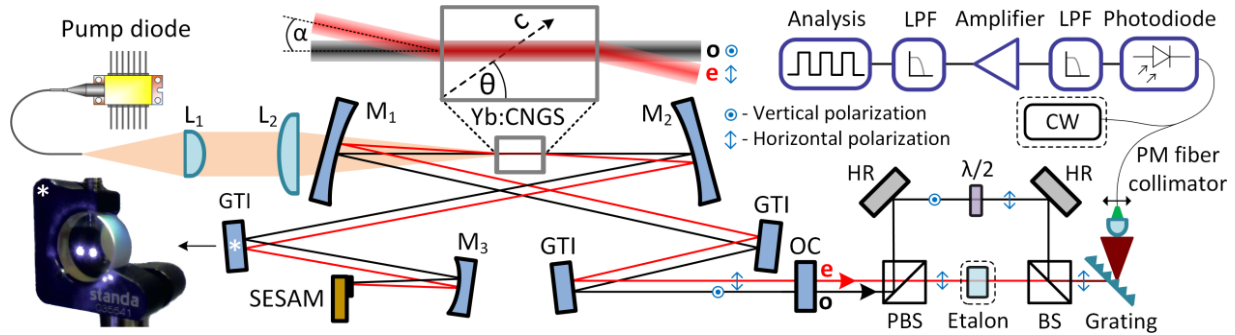

Fig. 1. Schematic of the dual-comb spectrometer based on the Yb:CNGS laser. Due to double refraction inside the birefringent gain medium, two intracavity beams are generated. The orthogonally polarized beams form an *o*-ray (black) and an *e*-ray (red) inside the crystal. The photograph (*) presents a real image of the two beams incident on one of the cavity mirrors. After being emitted from the oscillator, the beams are separated with a polarization beam-splitter (PBS). The polarization state of one of them is rotated with a half-wave plate (λ/2), while the other passes through an (optional) etalon. The beams are combined with a 50:50 beam-splitter (BS) and pass through a diffraction grating, which enables to select a narrowband component with a fiber collimator. The signal from a photodiode is low-pass-filtered (LPF) at 35 MHz and amplified before analysis. $L_1$ —18.4 mm aspheric lens; $L_2$ —100 mm spherical lens; $M_{1-3}$ — 100 mm concave spherical mirror; GTI — Gires–Tournois-Interferometer mirror; OC — output coupler; SESAM — semiconductor saturable absorber mirror; HR – highly-reflective silver mirror; CW – optional single-frequency laser used for relative linewidth measurement.

As mentioned earlier, our dual-comb laser concept relies on simultaneous generation of two spatially separated beams with orthogonal polarizations. This can be realized in a single cavity due to the double refraction phenomenon inside the birefringent Yb:CNGS gain medium. The crystal was cut according to a special scheme with an angle $\theta$ = 36° between optical axis *c* (oriented in the horizontal plane) and the normal propagation direction (see Fig. 1). Consequently, the beams polarized as an *o*- (vertical polarization) and a *e*-ray (horizontal) experience a non-uniform refraction angle. The resulting walk-off angle between the two Poynting vectors amounts to 2.3° for a center wavelength of 1050 nm and for the chosen crystal cut. In order to achieve simultaneous lasing at both polarizations (further termed as *o*- and *e*-beams), sufficient overlap between them and the pump beam must be present. This is possible if the AOI on the gain element is different for the two polarizations. Specifically, the AOI difference (angle $\alpha$ in Fig. 1) can be chosen so that it compensates for the spatial walk-off and allows collinear propagation of the beams inside the crystal. The situation depicted in Fig. 1 corresponds to a positive uniaxial crystal where the *e*-wave Poynting vector is shifted towards the optical axis (crystallographic *c*-axis) with respect to the wave vector. Since the two laser beams are spatially-separated outside the Yb:CNGS crystal (see picture in Fig. 1*), they also experience various reflection angles from the concave $M_{1,2}$ mirrors, which enables to align the cavity to be stable for both polarizations in the same time. Due to non-common intracavity paths and crystal dispersion, the mode-locked *o*/*e*-beams will exhibit different repetition rates, which is a prerequisite for DCS. Note that this simple polarization-multiplexing concept relies exclusively on the gain medium birefringence and it does not require any additional parts to be introduced. Moreover, all the cavity components together with the pump source are common for both beams, enabling to achieve superior dual-comb coherence compared with the systems sharing only a part of the laser resonator [25,27]. While in the described setup the $\Delta f_{rep}$ is fixed, its tunability can be easily realized e.g. by introducing an optical wedge in the intracavity path of only one of the laser beams.

In the laboratory routine, we initially aligned the cavity to achieve lasing of the *o*-beam, which was collinear with the pump and shared its polarization state. Minor fine-tuning of the SESAM and OC horizontal tilt enabled to initiate lasing at orthogonal polarization (*e*-beam) and allowed to control the energy balance between the two beams. The latter is essential for simultaneous dual-polarization output as it allows to balance the gain anisotropy of the uniaxial Yb:CNGS [32]. A subsequent mechanical perturbation of the SESAM initiated stable mode-locked dual-comb operation. The average output power of each beam was equal to 30 mW at an incident pump power of ~700 mW. Figure 2a shows the optical spectra measured with an optical spectrum analyzer (OSA; AQ6370, Yokogawa) and the corresponding second-harmonic intensity autocorrelation traces (APE PulseCheck). The emission spectrum of the *e*(*o*)-beam was centered at 1051 nm (1057 nm) with a 13.8 nm (13.4 nm) full-width-at-half-maximum (FWHM) bandwidth. The corresponding pulse duration amounted to 93 fs (88 fs), which results in a time-bandwidth product of 0.35 (0.32). Figure 2b presents the RF spectrum of the fundamental beat notes separated by $\Delta f_{rep}$ = 4.73 kHz and centered around 78.3 MHz (registered with FSW43, R&S). The insets show clean 1-GHz-spanning RF spectra of the individual isolated signals indicating undisturbed operation stability, which will be confirmed by additional measurements presented in the following section 3.

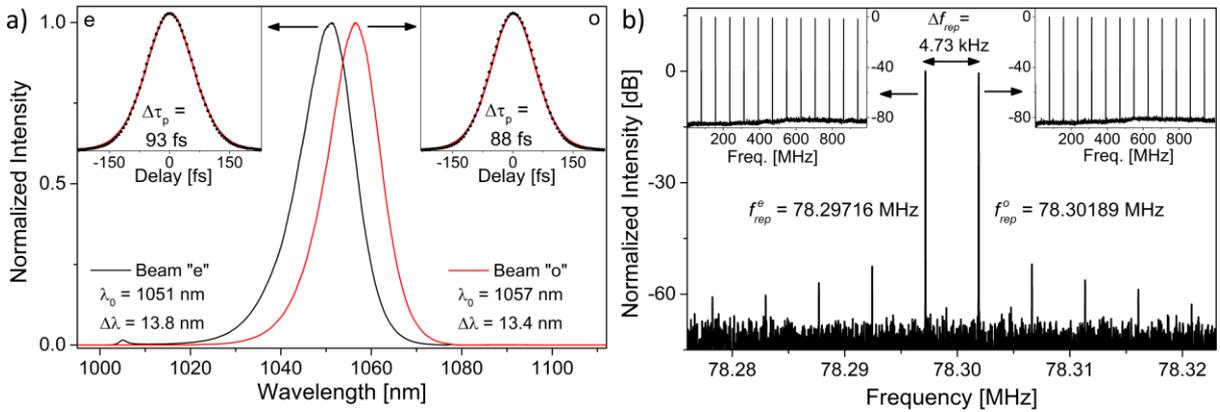

Figure 2. Performance of the dual-comb mode-locked Yb:CNGS laser. Independently normalized optical spectra of the *e*/*o*-beams; insets: corresponding autocorrelation traces with fits assuming sech$^2$-shaped pulses (a). RF spectrum of the fundamental beat notes of both beams around 78.3 MHz (resolution bandwidth: RBW = 100 kHz); insets: corresponding 1-GHz-wide RF spectra (RBW = 1 kHz) of the individual beams (b).

The measurements clearly indicate that the laser emitted two orthogonally-polarized, spatially-separated and parallel beams. In order to enable their mutual beating, we had to ensure their identical polarization states. Therefore, we separated the beams with a polarization beam-splitter (PBS) and rotated one of them by 90˚ with a half-wave plate ($\lambda$/2). The other arm of the interferometer contained an optional fused silica etalon with a 1 mm thickness. The beams were next combined with a 50:50 beam-splitter (BS) and passed through a transmission diffraction grating (1000 lines/mm), which enabled to select a narrowband ($\Delta\lambda$ = 0.8 nm) component with a PM fiber collimator. Spectral narrowing was crucial to avoid aliasing of the comb transferred to the RF domain. With a compression factor ($f_{rep}/\Delta f_{rep}$) of approximately 16500, the maximum optical bandwidth that can be mapped up to $f_{rep}$/2 amounts to 650 GHz (2.4 nm). By scanning the position of the collimator any available spectral region could be selected. The collected IR radiation was passed to a photodiode (FGA04, Thorlabs) and before analysis its electrical signal was subsequently amplified by a PE15A1012 (Pasternack) low-noise amplifier supplied with with home-made 35 MHz low-pass filters (LPF) both at its input and output.

## 3. Characterization of the free-running dual-comb laser

To prove the high mutual coherence between the generated combs and evaluate the feasibility of performing demanding mode-resolved dual-comb measurements, we characterized the laser for timing jitter and relative frequency fluctuations. We will start the discussion with the repetition rate stability analysis in the short- and long-term regime, followed by a focus on relative frequency fluctuations.

### 3.1 Repetition rate noise characteristics

To characterize the repetition rate phase noise of the two combs, we used a commercial phase noise analyzer (FSWP8, R&S). Both polarizations were measured independently in a frequency range from 1 Hz to 1 MHz. As shown in Fig. 3a, the system shows excellent phase noise performance. At 100 Hz of offset from the carrier located at ~78.3 MHz, the phase noise spectral density amounts to −100 dBc/Hz. Figure 3b shows the corresponding integrated timing jitter, which amounts to 14.8 ps, and 37.2 ps for the *e*- and *o*-beam, respectively. When compared with other reports on dual-comb lasers including this kind of characterization [19,27], the source presented in this work clearly exhibits superior stability. This can be mainly attributed to the different pump source employed. On the contrary to the powerful spatially-multimode diodes used in above-mentioned reports, we employed a single-mode laser diode with intrinsically low-noise operation, which further transfers to the phase fluctuations of the oscillator [30,31].

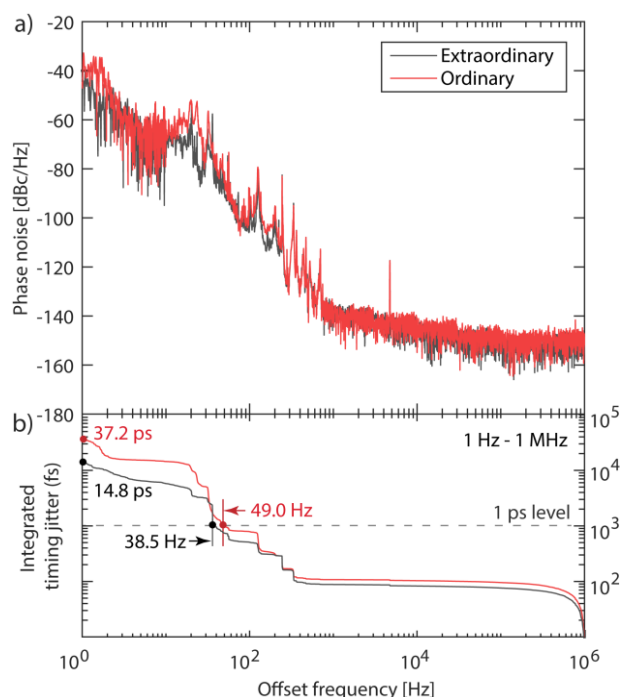

Figure 3. Repetition rate phase noise and timing jitter characterization. (a) Phase noise spectral density for the two repetition rates (*e*-/*o*-beams). (b) Corresponding integrated timing jitter.

In order to demonstrate the feasibility of prolonged dual-comb lasing and characterize its long-term stability, we recorded a spectrogram of the 100$^{th}$ harmonic of the repetition rates with a temporal resolution of 5 seconds, and a resolution bandwidth of 50 Hz. The spectra were next interpolated 16 times for accurate peak frequency estimation, and the harmonic frequency axis was normalized to the fundamental. Figure 4 plots the peak frequencies retrieved from the spectrogram recorded over an hour. Although the repetition rates drift during the measurement by ~275 Hz, the difference frequency oscillates by at most 3 Hz with 0.52 Hz standard deviation.

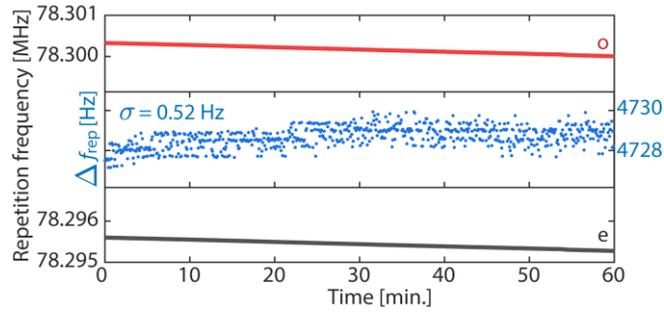

Figure 4. Long-term repetition rate stability characterization. Although the drift of the repetition rates is ~275 Hz within 60 minutes, the two frequencies closely follow each other down to a sub-hertz level – the standard deviation of the difference frequency is 0.52 Hz.

## 3.2 Relative-linewidth measurement

Our observation that the repetition rates precisely follow each other down to a sub-hertz level over an hour timescale does not necessarily imply suitability for mode-resolved free-running dual-comb spectroscopy. This is because the repetition rate characterization quantifies only the timing noise, while completely ignoring relative offset frequency fluctuations $\Delta f_{ceo}(t)$. From a practical standpoint, however, it is easier to measure the combined effect of the two unstable frequencies in an experiment with a CW single-mode laser, which is an established technique for characterizing single-cavity dual-comb lasers [7,8]. We heterodyned our dual-comb laser with a 1063.6 nm single-mode laser (D-C-1060, PicoQuant) and recorded the beating signal over 1 second. While this technique allows only for coarse estimation of the combs' optical linewidths (absolute stability), it is perfect for relative coherence assessments. This is because two individual beat notes are simultaneously recorded and their instantaneous difference frequency can be conveniently accessed.

Figure 5a plots the 1-s-long spectrogram of the beat notes with 30 μs temporal resolution (50 kHz RBW). One can observe frequency fluctuations with ~4.1 MHz peak-to-peak deviations, which correspond to an upper bound of the optical linewidth of ~3.1 MHz assuming uncorrelated frequency noise between the combs and the CW laser. More importantly, the instantaneous frequency of the two beat notes is visually similar (see zoomed panel), which is expected for high relative coherence. Using a digital difference frequency generation (DDFG) routine [33], we obtained the relative frequency fluctuations of the signal and characterized it in the frequency domain with a resolution bandwidth of 710 Hz, which corresponds to 10 repetition rate difference frequency periods (2.1 ms). This analysis follows techniques described in detail in our previous report [11]. The results of the relative coherence analysis are plotted in Fig. 5b-f. Firstly, despite the 4 MHz-wide fluctuations of the CW laser beat notes, the difference frequency (Fig. 5b) remains almost constant with a 30 kHz drift in total. This indicates a high degree of coherence between the two combs, as the drift within 1 second corresponds to a shift in DCS line position lower than 7 comb teeth (<500 MHz of optical resolution).

Fortunately, this number can be greatly improved if one uses frequency-tracking computational DCS phase correction algorithms developed recently [33–35]. A prerequisite for such algorithms to enable mode-resolved dual-comb measurements is the Nyquist criterion: the relative frequency between the two combs cannot fluctuate by more than half of the sampling interval equal to the repetition rate difference period [36], or equivalently the duration of the dual-comb IFG. In other words, as long as frequency-noisy dual-comb lines drift by less than half the spacing between the lines from IFG shot to shot, they can be digitally shifted back to their original position. Unfortunately, larger frequency excursions cannot be tracked because they become aliased – the frequency drift is much higher than accurately measurable during a single IFG period. Although an attempt to apply digital phase correction to such noisy systems will improve the look of the dual-comb spectrum and restore its discrete character with sharp lines, optical sample information will leak from a single comb tooth into its neighbors, thus making spectroscopic data inaccurate. This is because the digital algorithm will be blind to drifts exceeding multiple line spacings (shot-to-shot) and will simply lock the position of the drifted lines closest to those already averaged. In practice, a maximally shared laser cavity design along with low-noise pump sources renders relative linewidths compatible with the correctability criterion. Even the most demanding free-running DCS are possible, where a drift by more than one line spacing would have severely affected absorption lineshapes [11,12]. Still, one has to keep in mind that the resolution in such systems always remains limited by the comb optical linewidth.

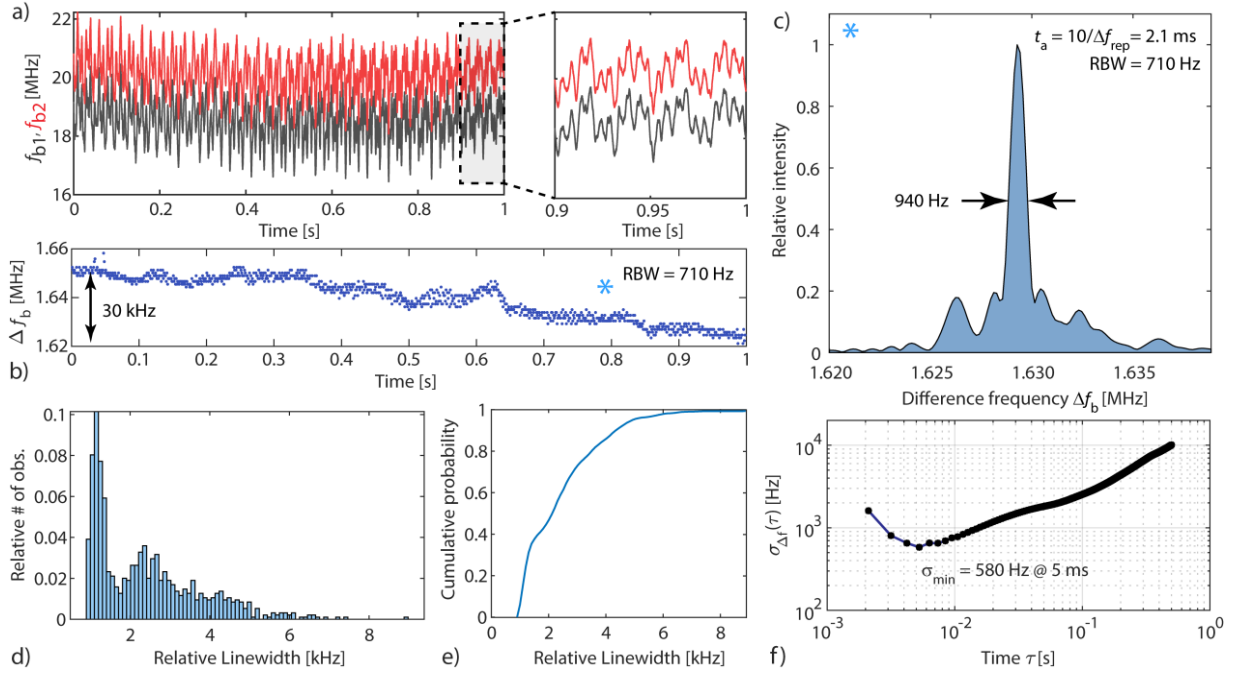

Figure 5. Beat with CW laser. (a) Spectrogram of the beating signal with 30 μs temporal resolution. (b) Difference frequency of the beat notes in (a). (c) Example difference frequency beat-note at 0.8 s. (d) Histogram of the relative linewidth. (e) Cumulative probability density function retrieved from (d). (f) Allan deviation of the difference frequency from (b). The global minimum is achieved at 5 ms, followed by drift at longer time scales.

In our system, the required shot-to-shot relative stability for mode-resolved DCS must be lower than ~2.3 kHz over 1/4.73 kHz = 211 μs time scale. To characterize it, we measured the full-width-at-half-maximum (FWHM) of the relative beat note for all time instances of the data from Fig. 5b. Overall, the system shows relative beat notes as narrow as ~940 Hz over 10 periods of the IFG equal to 2.1 ms (Fig. 5c). Nevertheless, rather than focusing on the best case, we performed a statistical linewidth analysis as depicted in Fig. 5d,e. The mode of the distribution is 1.2 kHz, while the mean is 2.4 kHz. The empirical cumulative distribution function (Fig. 5e) indicates that linewidths narrower than 5 kHz are obtained with 95% probability. Given that this level of fluctuations occurs over timescales 10× longer than needed for mode-resolved operation, we are confident that the shot-to-shot stability fully complies with requirements for computational frequency tracking. It should be also noted that without computational assistance, the system will be still compatible with mode-resolved operation up to 5 ms, as concluded from the global minimum of the difference frequency Allan deviation (Fig. 5f).

## 4. Dual-comb spectroscopy

To validate the results of our stability analysis and prove the spectroscopic capabilities of the dual-comb source, we performed a free-running DCS experiment. Rather than acquiring the DCS IFG with an oscilloscope, we utilized a 16-bit quadrature demodulator (FSW43, R&S), recording complex IFG data centered around 19 MHz for 1 second. As a test medium, we measured a 1-mm thick fused-silica etalon with a dielectric coating increasing its reflectance around 1 μm. The sample was inserted into one of the dual-comb spectrometer arms (*e*; see Fig. 1). For spectral normalization purposes, an analogous DCS signal was recorded without the etalon referred to as a reference measurement. Both signals were computationally phase-corrected and coherently averaged. Before we proceed to the recovered transmittance spectra, we focus first on the characteristics of the dual-comb signals shown in Fig. 6.

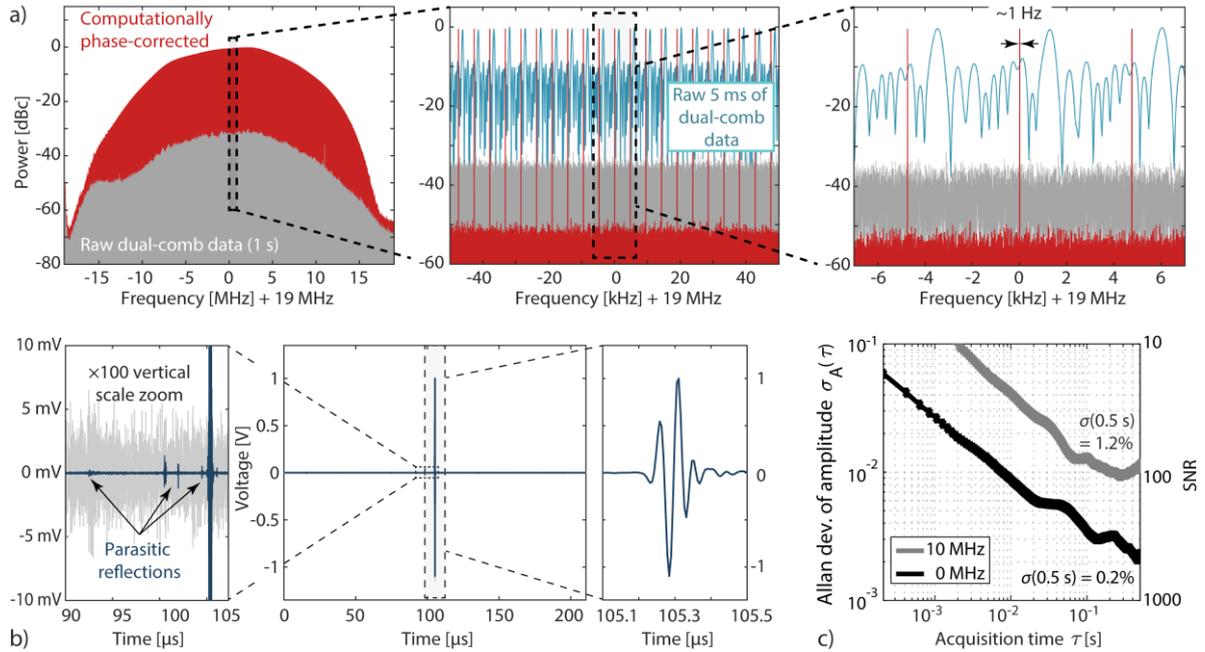

Figure 6. Dual-comb signal analysis. (a) Raw and computationally phase-corrected dual-comb spectra. Zoomed panels show a comparison of the corrected spectrum with original unprocessed 1-s-long data (grey) and cropped 5-ms-long acquisition (blue). After the computational correction, the RF comb linewidth reaches an acquisition-time-limited linewidth (anywhere in the spectrum, not only around the center). Without digital enhancement, the system is capable of tooth-resolved measurements over 5 milliseconds with ~10 dB carrier-to-noise ratio. (b) Coherently averaged 4742 dual-comb IFGs (without etalon) plotted along with a single-shot unprocessed trace. The traces are normalized to 1 V. The averaging gain reaches 99.4% of the theoretical limit (36.76 dB). (c) Amplitude Allan deviation analysis of the beat notes located close (0 MHz), and far-away (10 MHz) from the carrier frequency of 19 MHz.

As expected from the stability analysis, the high mutual coherence allows to perform tooth-resolved measurements over ~5 milliseconds with ~10 dB carrier-to-noise ratio and percentage precision. Even though over extended time scales exceeding 1 second (Fig. 6a), the DCS lines lose their discrete character, the sufficient mutual coherence between the combs permits to faithfully track the phase of the IFG and its time-varying duration. We have already demonstrated the phase correction algorithm in demanding high-resolution Doppler-limited spectroscopy applications [11], and here it is also used to unlock the full potential of the system. After digital correction, a boost in the spectral SNR exceeds 36 dB, which agrees well with expectations from the number of averaged frames equal to 4742. The dual-comb spectral linewidth is limited by the acquisition time to ~1 Hz. Nevertheless, it should be noted that the algorithm corrects only for RF signal fluctuations, while leaving optical frequency fluctuations intact. In other words, the obtainable resolution remains limited by the optical linewidth of the comb on the order of megahertz.

Fig. 6b shows a comparison of the raw (single-shot) and coherently averaged IFGs. Because the original data are complex and carrier-free (due to complex demodulation), for visualization purposes we frequency-shifted the complex signals to show analogous baseband real IFGs. Raw single-shot (uncorrected) IFGs are spurious-free and already show a relatively high dynamic range of 52 dB, which improves to 89 dB upon averaging. Therefore, the difference between the two traces becomes noticeable only at high vertical scale magnification. Prolonged averaging reveals the existence of very weak bursts surrounding the IFG center-burst, all attributable to parasitic reflections at imperfect AR-coatings on the crystal. The obtainable precision with the system in one second of averaging characterized using Allan-Werle deviation analysis [37] is 0.2% for lines around the carrier frequency (strongest), degrading to 1.2% for lines 10 MHz away from the carrier (Fig. 6c). The corresponding spectroscopic SNRs are 500 and 83, respectively.

In the final step, we compared the digitally corrected dual-comb spectra (etalon and reference) with a simultaneous measurement using an optical spectrum analyzer (OSA, AQ6370B Yokogawa), which was used for optical frequency axis calibration. Unlike in previous demonstrations, we performed a measurement in both amplitude and phase. The latter was possible by forcing the carrier phase of the two measurements to be identical. Consequently, the measurement may in general possess a constant phase offset compared to fully CEO-phase-stabilized systems. Fortunately, for many applications it is the shape of the dispersion spectrum than matters. Overall, excellent agreement is obtained between the DCS and OSA measurement plotted in Fig. 7. Slight deviations between the curves occur around the peaks, which may be related to photodetector nonlinearities in DCS and a finite resolution of the OSA (0.01 nm).

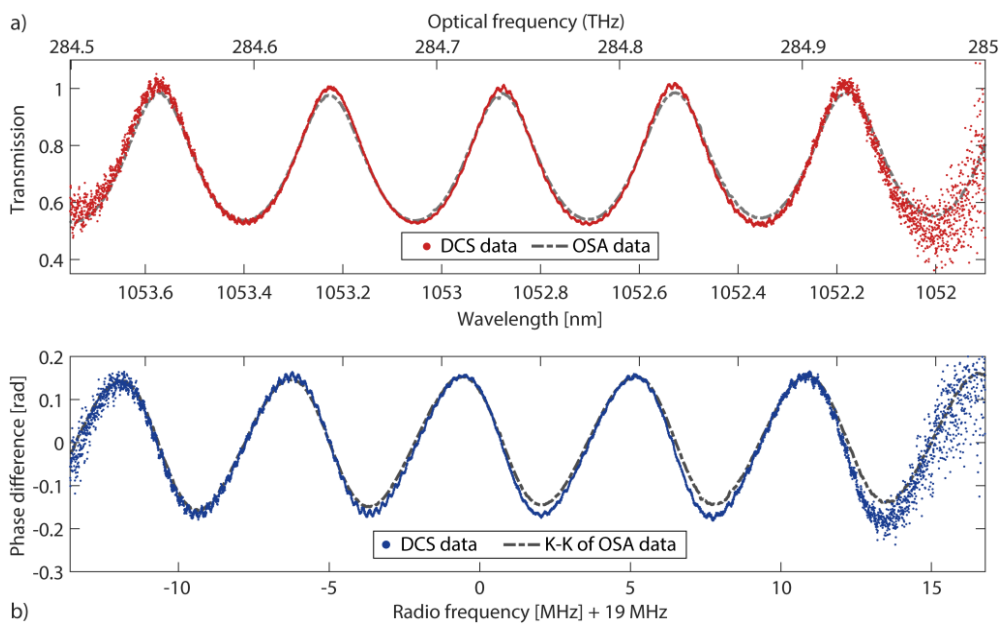

Figure 7. Dual-comb spectroscopy of a fused silica etalon compared with an optical spectrum analyzer (OSA) measurement. (a) Optical transmission. (b) Optical phase difference. The expected optical phase (dashed line) was derived from the OSA transmission using the Kramers-Kronig (K-K) relation.

As expected from the Allan analysis, the spectroscopic precision in the central part of the spectrum is higher than on the wings, which directly relates to a steep beat note power roll-off above 10 MHz from the spectral center. The etalon effect visible in the transmission and dispersion spectrum stems from the aforementioned imperfections of the crystal AR-coatings, as no interferogram windowing was used to improve the spectroscopic performance at the expense of loss in resolution. The DCS spectral coverage reaches ~500 GHz (~1.8 nm around 1052.8 nm), which is merely a limitation of the non-aliased RF bandwidth. The employed grating-based tunable optical filter allows for selection of an arbitrary center wavelength on demand within an available spectral range.

## 5. Conclusions

We have demonstrated a novel scheme for dual-comb generation from a single-cavity mode-locked laser. It is based on intrinsic polarization-multiplexing originating from double refraction inside a birefringent gain medium and hence does not require any additional components to be introduced to a standard laser resonator. Despite its simplicity, all the cavity components are common for both beams, enabling to achieve high relative coherence of the free-running combs, a key-aspect for DCS.

The proposed concept was validated in the Yb:CNGS laser generating two orthogonally-polarized sub-100 fs pulse trains at the central emission wavelength of 1.05 µm and with repetition rate offset of 4.73 kHz (at 78.3 MHz). Note that the femtosecond pulse duration is crucial regarding potential transfer of the generated dual-comb radiation to other spectral regions via nonlinear frequency conversion. The developed source was precisely characterized for timing jitter and relative frequency fluctuations. A detailed investigation revealed unprecedented stability with sub-kilohertz relative linewidth over 10 IFG periods (2.1 ms) permitting mode-resolved measurements over millisecond time scales. Moreover, the measurement time can be greatly extended by using computational phase correction exploiting the high mutual coherence between the combs. We unveiled the potential of the developed system by presenting mode-resolved transmission and dispersion measurements of a low finesse etalon with permille amplitude precision obtainable within 1 second. To the best of our knowledge, our source exhibits the lowest relative noise ever demonstrated for a solid-state dual-comb laser, which may be attributed to its simple single-cavity design and the stability of the employed pump source.

What is particularly important regarding spectroscopic applications is that our straightforward concept can be in principle applied to any bulk laser based on an optically anisotropic gain crystal. Consequently, it can be exploited to realize dual-comb laser sources operating in mid-IR spectral band, a region containing the majority of characteristic ro-vibrational molecular transitions. This can be achieved by developing high-power, femtosecond 2 µm lasers based on $Tm^{3+}$- [38], $Ho^{3+}$- [39] or $Cr^{2+}$-doped [40,41] crystals and its subsequent nonlinear conversion to the longer wavelength range [42] or by direct dual-comb generation from $Fe^{2+}$ lasers up to 6 µm [40,43]. Consequently, the reported technique paves the way for a new class of single-cavity dual-comb lasers, which, due to their low-noise free-running operation, will enable extremely precise spectroscopic measurements without any active stabilization in the near future.

**Acknowledgments**

We would like to acknowledge dr Marcin Motyka from Laboratory for Optical Spectroscopy of Nanostructures, Wrocław University of Science and Technology for lending us a single-frequency CW laser and R&S Poland for lending us the phase noise analyzer.

**Funding**

National Science Centre (NCN, Poland); grant no. 2015/18/E/ST7/00296.